\documentclass[aps,prb,twocolumn,floatfix,superscriptaddress,showpacs,footinbib]{revtex4}
\usepackage{amsmath,amssymb,bm,graphicx,url,epsfig,bbm}
\usepackage{ulem}
\usepackage[ansinew]{inputenc}

\usepackage{color}

\begin{document}

\title{Transport through a quantum dot with excitonic dot-lead coupling} 

\author{Florian Elste}
\affiliation{Department of Physics, Columbia University, 538 West 120th Street, New York, NY 10027, USA}

\author{David R. Reichman}
\affiliation{Department of Chemistry, Columbia University, 3000 Broadway, New York, NY 10027, USA}

\author{Andrew J. Millis}
\affiliation{Department of Physics, Columbia University, 538 West 120th Street, New York, NY 10027, USA}

\date{\today}

\begin{abstract}
We study the effect of a dot-lead interaction on transport through a quantum dot hybridized to two semi-infinite Luttinger-liquid leads. A bosonization approach is applied to treat the interaction between charge fluctuations on the dot and the dynamically generated image charge in the leads. The nonequilibrium distribution function of the dot and the tunneling current are computed within a master-equation approach. The presence of the excitonic dot-lead coupling is found to enhance transport in the vicinity of the Coulomb-blockade threshold. This behavior is in contrast to the usual power-law suppression of electronic tunneling which is found if this interaction is ignored.
\end{abstract}


\maketitle

\section{Introduction}

Understanding nonequilibrium phenomena in transport through quantum dots and single-molecule devices is of fundamental interest and is a major challenge in the field of nanoscience. Much attention has been paid to the interplay of the dot-lead hybridization and on-dot interactions involving vibrational degrees of freedom or the local repulsions which lead to magnetic moments and to Kondo physics. However, the effect of an `excitonic' interaction between the charge on the dot and the charge on the leads has been studied little, even though it clearly should be present on physical grounds. 
In this paper we address this physics in the context of a quantum dot connected to two one-dimensional leads. The model we consider is inspired by recent experimental investigations of the conductance of a molecule placed in the gap created by breaking a carbon nanotube,\cite{Guo06,Chandra09} although theory and experiment are not yet at the point where a quantitative comparison can be made. 

One may view a quantum dot with one-dimensional leads as an impurity embedded in a Luttinger liquid. Fabrizio \textit{et al.}\cite{Fabrizio94} and Maurey and Giamarchi\cite{Maurey} have studied the case of an impurity described as a short-ranged potential scatterer without dynamical charge fluctuations. Lerner \textit{et al.}\cite{Lerner} have taken into account finite dot-lead hybridization but have not considered a Coulombic dot-lead coupling. Goldstein \textit{et al.}\cite{Goldstein} have investigated the effect of a Coulombic dot-lead interaction on the dynamics of the population of a quantum dot using density matrix renormalization group and classical Monte Carlo simulations. 

In a previous paper we studied the effect of a coupling between dot and lead charge densities on the relaxational dynamics of a quantum dot side-hybridized to a Luttinger-liquid lead.\cite{Elste} We found that this `excitonic' coupling has important consequences for electronic correlations and may enhance the tunneling of electrons in the regime of weak hybridization. The present paper extends our analysis to the case of a quantum dot placed between two leads such that it \textit{cuts} the Luttinger liquid into two semi-infinite quantum wires. This situation differs in several respects from that considered in our previous work. Most fundamentally, intrinsically nonequilibrium behavior driven by a current flow across the dot is possible. Further, because we deal with semi-infinite systems, boundary exponents appear instead of bulk exponents. Finally, the presence of two leads means that tunneling into one lead is modified by orthogonality effects arising from the excitonic coupling to the other lead. These differences turn out to produce significant changes in the results.

The rest of the paper has the following structure. In Sec.~\ref{Model} we introduce the model. In Sec.~\ref{Bosonization} we explain the bosonization scheme which we use. In essence the idea is to write the model in a basis corresponding to a translation-invariant system and to note that the breaking of translational invariance corresponds to imposing boundary conditions on the lead wavefunction which may be satisfied by choosing particular linear combinations of wavefunctions in a way inspired by the "method of images" in electrostatics.\cite{Eggert95,Fabrizio95,Eggert96,Giamarchi04} Section~\ref{can_trans} solves the resulting the model by canonical transformations \`{a} la Schotte and Schotte.\cite{Schotte69} Results for the tunneling rates and current-voltage characteristics, calculated to leading nontrivial order in the dot-lead hybridization, are presented in Sec.~\ref{Electroniccorrelations} and Sec.~\ref{Currentvoltagecharacteristics}. Section~\ref{Conclusions} presents a summary and conclusions, and indicates possible directions for future research. 

\section{Model \label{Model}}

Our system consists of leads, a quantum dot, and hybridization and excitonic coupling terms. It is described by a Hamiltonian of the form 
\begin{equation}
H = H_\text{lead} + H_\text{dot} + H_\text{exc} + H_\text{hyb}.
\label{H}
\end{equation}
For simplicity, we take the quantum dot to have one (spin-degenerate) level and a repulsive interaction. However, we emphasize that the method is general and would work in more complicated situations. The quantum-dot Hamiltonian $H_\text{dot}$ is thus
\begin{equation}
H_\text{dot}= \varepsilon_d \, n_d +\frac{U}{2} n_d(n_d-1).
\label{Hdot}
\end{equation}
Here $U$ is the local Coulomb repulsion, $n_d=\sum_{\sigma} d_{\sigma}^\dagger d_{\sigma}$ is the operator giving the total number of electrons on the dot and $d_{\sigma}^\dagger$ creates an electron with energy $\varepsilon_d$ and spin $\sigma$ on the dot.

We model the lead part of the Hamiltonian as two disconnected half-chains. Motivated by the possibility of nanotube leads we allow the two leads denoted by $\alpha=L,R$ to have an orbital (channel) degeneracy labeled by $1\leq a \leq N$ as well as a spin degeneracy. The lead Hamiltonian is then
\begin{align}
H_\text{lead} &~=~ \sum_{\alpha a k \sigma} \epsilon_{\alpha k} c^{\dagger}_{\alpha a k \sigma} c_{\alpha a k \sigma} +H_\text{int}.
\label{Hlead}
\end{align}
Here $c^{\dagger}_{\alpha a k \sigma}$ creates an electron in a state of energy $\epsilon_{\alpha k}$ in orbital $a$ of lead $\alpha$, and $H_\text{int}$ labels the interactions among lead electrons, which will be discussed in more detail below. 

\begin{figure}[!t]
\begin{center}
\includegraphics[width=7.5cm,angle=0]{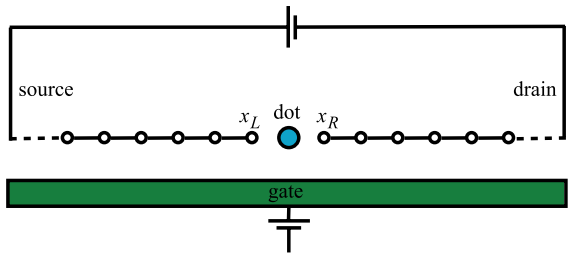}
\caption{Schematic of a one-dimensional lead-dot-lead system showing two leads, possibly maintained at a nonvanishing voltage difference, the quantum dot (filled circle, green online) and a gate which screens the long-ranged part of the Coulomb interaction.}\label{model_fig} 
\end{center}
\end{figure}

The situation we consider is sketched in Fig.~\ref{model_fig}.
The coordinate system is defined such that the quantum dot is at $x=0$ and lead $\alpha$ terminates at $x=x_\alpha$ with $x_L<0$ and $x_R>0$. We assume that in the absence of the dot the leads are decoupled so that the electronic boundary condition is 
\begin{equation}\label{constraint}
\psi_{\alpha a \sigma}(x_\alpha)=0,
\end{equation}
where $\psi_{\alpha a \sigma}(x)$ creates a lead electron at position $x$. 

We restrict ourselves to the case of short-ranged interactions between the conduction electrons, assuming in particular that the interaction between electrons in different leads is negligible. The effect of interactions between leads is an important open question.
Interactions become short-ranged in the presence of a metallic gate, which screens the long-ranged part of the Coulomb force. We shall need an explicit form for the charge-channel interaction in the leads, which we take to be 
\begin{equation}
H_\text{int}=\frac{1}{2}\sum_{\alpha=L,R} V^c_\alpha \int^\alpha dx \left[ \rho_\alpha^\text{tot} (x)\right]^2 + \dots,
\label{Hcharge}
\end{equation}
where $\int^\alpha$ indicates an integral over the region of lead $\alpha$ and
\begin{equation}
\rho_\alpha^\text{tot}(x)=\sum_{a\sigma} \rho_{\alpha a \sigma}(x)
\label{rhotot}
\end{equation} 
with 
\begin{equation}
\rho_{\alpha a \sigma}(x) = \sum_{qk} e^{iqx} c^\dagger_{\alpha a (k+q) \sigma} c_{\alpha a k \sigma}
\end{equation} 
denotes the total charge density (summed over all channels and spins). 
We shall not need to specify other potential contributions to $H_\text{int}$.

We include two forms of dot-lead couplings. The first is the standard hybridization
\begin{equation}
H_\text{mix} = \sum_{\alpha}\int^\alpha dx \sum_{a \sigma} \left[ \mathcal{T}_\alpha(x) \, d^{\dagger}_{\sigma} \psi_{\alpha a \sigma}(x) + \text{h.c.} \right],
\label{Hmix}
\end{equation}
with the tunneling amplitudes $\mathcal{T}_\alpha(x)$ strongly peaked at $x$ in the vicinity of the lead edges. The second term is a (screened) Coulombic dot-lead coupling, 
\begin{equation}
H_\text{exc} = n_d \sum_{\alpha}\int^\alpha dx \, W_{\alpha}(x) {\rho}_{\alpha }^\text{tot}(x),
\label{Hcoul}
\end{equation}
where the dot-lead interaction $W_{\alpha}(x)$ is also peaked for $x$ near the lead edge.

\section{Bosonization scheme}\label{Bosonization}
\subsection{General formalism}

Because the leads are one-dimensional and the dispersion may be linearized near the Fermi level, we expect to be able to represent the low-energy degrees of freedom in terms of bosons representing propagating charge and spin fluctuations.\cite{Luther,Haldane81} The semi-infinite nature of the leads presents a technical difficulty in that the standard bosonization formulas are derived for translation-invariant systems, whereas in the model of interest translation invariance is strongly broken. To deal with this situation we use ideas introduced by Kane and Fisher,\cite{Kane92} Eggert and Affleck\cite{Eggert95} and Fabrizio and Gogolin.\cite{Fabrizio95} For each lead we introduce a translation-invariant reference system [Eq.~(\ref{Hlead}) with the range of $x$ extended from $-\infty$ to $+\infty$], which we bosonize as usual. We then impose a boundary condition which selects from the reference problem only those states which fulfill the physical boundary condition.

\subsection{Bosonization of reference problem}
We now recall the bosonization of the translation-invariant reference problem, in order to define notation. We combine spin and orbital quantum numbers into a superindex $\beta=1,\dots,M$ and introduce the operators $\rho^\lambda_{\alpha\beta}(x)$ describing right ($\lambda=+$) and left ($\lambda=-$) moving particle-hole pairs in orbital $\beta$ of the reference system corresponding to lead $\alpha$. From these we define boson operators $\phi_{\alpha,\beta}^0$, $\theta_{\alpha,\beta}^0$ via
\begin{align}
\nabla {\phi}^0_{\alpha\beta}(x) &~=~ \left[ \rho^+_{\alpha\beta}(x) + \rho^-_{\alpha\beta}(x)\right], \\
\nabla {\theta}^0_{\alpha\beta}(x) &~=~ - \left[ \rho^+_{\alpha\beta}(x) - \rho^-_{\alpha\beta}(x) \right]
\end{align}
that obey the volume commutation relation
\begin{equation}\label{commutation}
\left[\phi^0_{\alpha\beta}(x),\nabla \theta^0_{\alpha'\beta'}(x')\right]=\frac{i}{\pi} \, 
\delta_{\alpha\alpha'}
\delta_{\beta\beta'}
\delta(x-x')
\end{equation} 
and are related to the total particle density in state $\beta$ by
\begin{equation}
\rho_{\alpha\beta}(x)=\nabla \phi^0_{\alpha\beta}(x).
\label{rhodef}
\end{equation}
Expressing the lead electron creation operator 
\begin{equation}
\psi_{\alpha\beta}(x)=\sum_{\lambda=\pm}\psi_{\lambda\alpha\beta}(x)
\label{psiLR}
\end{equation} 
in terms of these bosons fields gives
\begin{equation}
\psi_{\lambda \alpha\beta}(x)=\frac{U_{\lambda\alpha}}{\sqrt{2\pi\eta}}e^{i\lambda k_F x}e^{ i \pi \left[ \lambda{\phi}^0_{\alpha\beta}(x)- \theta_{\alpha\beta}^0(x) \right]}.
\label{psi}
\end{equation}
Here the small positive infinitesimal factor $\eta$ arises from the correct normal ordering of the operators \cite{Haldane81}.
The Klein factor $U_{\alpha}$ carries the Fermi statistics and a time dependence related to the lead chemical potential 
$\mu_\alpha$,
\begin{equation}
U_{\lambda \alpha}(t) = e^{i\mu_\alpha t} U_{\lambda \alpha}(0).
\end{equation}

Under reasonable conditions (no Umklapp scattering, standard interactions) the low-energy part of the Hamiltonian for the reference translation-invariant problem may be diagonalized in terms of boson fields $\phi_{\alpha b},\theta_{\alpha b}$ which are linearly related to the original operators $\phi_{\alpha\beta}^0,\theta_{\alpha\beta}^0$ and obey the same commutation relations, Eq.~(\ref{commutation}).~\cite{Luther,Haldane81} 
It is convenient to rearrange the $\theta_{\alpha b}$, $\phi_{\alpha b}$ into {\it chiral} fields ${\tilde \phi}_{\alpha b}^\pm$ conventionally defined by
\begin{equation}
\tilde{\phi}_{\alpha b}^\pm(x) \equiv \tilde{\theta}_{\alpha b}(x) \pm \tilde{\phi}_{\alpha b}(x), \label{chiral} 
\end{equation}
in terms of which the reference Hamiltonian becomes
\begin{equation}\label{Hamiltonian3}
H_\text{ref} = \sum_{\alpha b} \frac{\pi v_{\alpha b}}{4} \int dx \left( [\nabla {\tilde \phi}^+_{\alpha b}(x)]^2 + [\nabla {\tilde \phi}^-_{\alpha b}(x)]^2 \right)
\end{equation}
with the velocity parameters $v_{\alpha b}$ and coefficients of the linear transformation relating the ${\tilde \phi}_{\alpha b}^\pm$ to the $\phi_{\alpha \beta}^0$, $\theta_{\alpha \beta}^0$ determined by the bare velocities and interactions of the lead eigenstates. Note that $[{\tilde \phi}_{\alpha b}^+,{\tilde \phi}_{\alpha' b'}^-]=0$ while 
\begin{equation}\left[{\tilde \phi}_{\alpha b}^\pm(x),\nabla{\tilde \phi}_{\alpha' b'}^\pm(y)\right]=\pm\frac{2i}{\pi}\delta_{\alpha\alpha{'}}\delta_{bb{'}}\delta(x-y),
\label{chiralcr}
\end{equation}
which among other things implies that $\tilde{\phi}_{\alpha b}^\pm(x,t)$ are respectively functions of $x\pm v_F t$ only.\cite{Giamarchi04}

In the general case the transformations are complicated, but in the physically most relevant case where all channels in a given lead have the same bare velocity and the interactions conserve the total lead density, two simplifications occur. First, the transformation does not mix the $\theta$ and $\phi$ variables so that
\begin{eqnarray}
\phi^0_{\alpha\beta}&=&\frac{1}{2}\sum_b {\mathcal C}^\alpha_{\beta b}\left({\tilde \phi}^+_{\alpha b} -{\tilde \phi}^-_{\alpha b} \right),
\label{Cdef}
\\
\theta^0_{\alpha\beta}&=&\frac{1}{2}\sum_b \left({\mathcal C^\alpha}\right)^{-1}_{b\beta}\left({\tilde \phi}^+_{\alpha b} + {\tilde \phi}^-_{\alpha b} \right).
\label{Cinvdef}
\end{eqnarray}
Second, one channel (which we take to be $b=1$) describes the total charge density in lead $\alpha$ so that $\nabla \phi_{\alpha, b=1}$ is linearly proportional to $\rho^\text{tot}_\alpha$ [Eq.~(\ref{rhotot})]. Specifically,
\begin{eqnarray}\label{basis1}
\frac{1}{2}\left({\tilde \phi}^+_{\alpha,b=1} -{\tilde \phi}^-_{\alpha,b=1} \right)
 &=&\sqrt{ \frac{1}{K_cM}}\sum_{\beta}\phi^0_{\alpha\beta}
\\ 
\frac{1}{2}\left({\tilde \phi}^+_{\alpha,b=1} +{\tilde \phi}^-_{\alpha,b=1} \right) &=& \sqrt{\frac{K_c}{M}}\sum_{\beta}\theta^0_{\alpha\beta}
\end{eqnarray}
and, conversely, 
\begin{eqnarray}
\phi^0_{\alpha\beta} &=&\frac{1}{2} \sqrt{\frac{K_c}{M}}\left({\tilde \phi}^+_{\alpha,b=1} -{\tilde \phi}^-_{\alpha,b=1} \right)+\dots,
\label{basis2phi}\\
\theta^0_{\alpha\beta} &=&\frac{1}{2} \sqrt{\frac{1}{K_cM}}\left({\tilde \phi}^+_{\alpha,b=1} +{\tilde \phi}^-_{\alpha,b=1} \right)+\dots,
\label{basis2theta}
\end{eqnarray}
with the ellipses representing the other terms ${\tilde \phi}_{\alpha,b \geq 2}^+\pm{\tilde \phi}_{\alpha,b \geq 2}^-$ needed to make up the full operator.

The `Luttinger parameter' $K_c$ and charge-channel velocity $v_{\alpha, b=1}\equiv v_c$ are related to the bare charge-channel interaction $V^c_\alpha$, Eq.~(\ref{Hcharge}), and the bare Fermi velocity $v_F$ by 
\begin{equation}
K_{c} = \frac{1}{\sqrt{1+\frac{MV^c_{\alpha}}{\pi v_F}}}, \quad v_{c} = v_F\sqrt{1+\frac{MV^c_{\alpha}}{\pi v_F}}.
\end{equation}
Transcribing Eq.~(\ref{psi}) into the new basis yields
\begin{align}\label{psidef1}
\psi_{\lambda\alpha\beta}& (x) ~=~ U_{\lambda\alpha} \, e^{i\lambda k_Fx} \, \psi_{\lambda \alpha\beta}^\text{rest}(x) \nonumber \\
\times \, & e^{i\frac{\pi}{\sqrt{M}} \left[\left( \lambda \frac{\sqrt{K_c}}{2}-\frac{1}{2\sqrt{K_c}}\right){\tilde \phi}^+_{\alpha,b=1}(x)-\left( \lambda \frac{\sqrt{K_c}}{2}+\frac{1}{2\sqrt{K_c}}\right){\tilde \phi}^-_{\alpha,b=1}(x)\right]} 
\end{align}
with $\psi_{\lambda \alpha\beta}^\text{rest}(x)$ an exponential of a combination of the ${\tilde \phi}^\pm_{\alpha b}$ with $b=2,\dots,M$.

\subsection{Implementation of boundary conditions}

Equation (\ref{psiLR}), in combination with the constraint, Eq.~(\ref{constraint}), on the lead wave function implies 
\begin{equation}\label{condition}
 U_{+,\alpha}e^{i k_F x_\alpha} \ e^{ i \pi {\phi}^0_{\alpha\beta}(x_\alpha,t)} + U_{-,\alpha}
 e^{-i k_F x_\alpha} \ e^{ -i \pi {\phi}^0_{\alpha\beta}(x_\alpha,t)} = 0.
\end{equation}
for all $\beta$ and $\alpha$.
Equation (\ref{condition}) is seen to be satisfied if we impose the two conditions
\begin{equation}
\text{(i)}~\cos \left[ \pi \phi^0_{\alpha\beta}(x_\alpha,t) +k_F x_\alpha \right] = 0, \quad \text{(ii)} \ U_{+,\alpha}=U_{-,\alpha}.
\end{equation}

Condition (ii) is the statement that perfect reflection at the channel edge means that the Klein factor, which carries the position dependence only via the chemical potential, is the same for left and right movers in each lead $\alpha$
so that henceforth we drop the subscript $\lambda$ in $U_{\lambda \alpha}$.

Condition (i), in combination with Eqs.~(\ref{basis2phi})---(\ref{basis2theta}) says that at the lead boundaries $x=x_\alpha$ the difference between the $+$ and $-$ chiral fields must be time independent. Shifting the origin of the coordinates for lead $\alpha$ to $x_\alpha$ we find
\begin{equation}\label{constraintXX}
\tilde{\phi}_{\alpha b}^+(0,t) - \tilde{\phi}_{\alpha b}^-(0,t) =D_{\alpha b}
\end{equation}
with the constants $D_{\alpha b}$ such that [using Eq.~(\ref{Cdef})] 
\begin{equation}\label{Db_eq}
\left( 1-\frac{2k_Fx_\alpha}{\pi}\right)=\sum_bC^\alpha_{\beta b}D_{\alpha b}.
\end{equation}
Because the chiral fields $\tilde{\phi}_{\alpha b}^\pm(x,t)$ are respectively functions of $x\pm v_F t$ only we can extend Eq.~(\ref{constraintXX}) to all space and time as
\begin{equation}\label{constraintYY}
\tilde{\phi}_{\alpha b}^+(x,t) - \tilde{\phi}_{\alpha b}^-(-x,t) = D_{\alpha b}.
\end{equation}
Inverting Eq.~(\ref{Db_eq}), noting that the left-hand side is independent of $\beta$ and using the fact that the charge channel is the same in original and eigenstate variables we find
\begin{equation}
D_{\alpha,1}=\sqrt{\frac{M}{K_c}}\left( 1-\frac{2k_Fx_\alpha}{\pi}\right), \quad D_{\alpha,b\neq 1}=0.
\end{equation}

To compute correlation functions for operators corresponding to the semi-infinite Luttinger liquid one rewrites a general expression 
in terms of the chiral operators ${\tilde \phi}_{\alpha b}^\pm$, uses the boundary condition Eq.~(\ref{constraintXX}) to eliminate, say, ${\tilde \phi}_{\alpha b}^-(x,t)$ in terms of ${\tilde \phi}_{\alpha b}^+(-x,t)$ and $D_{\alpha b}$, and then computes the ${\tilde \phi}_{\alpha b}^+$ correlation functions in the usual way from the ${\tilde \phi}_{\alpha b}^+$ part of Eq.~(\ref{Hamiltonian3}). For example, the fermion operator becomes
\begin{widetext}
\begin{equation}
\psi_{\lambda\alpha\beta}(x) 
= U_{\lambda\alpha} \, e^{i\lambda k_Fx} \, e^{i\frac{\pi}{\sqrt{M}}\left( \lambda \frac{\sqrt{K_c}}{2}+\frac{1}{2\sqrt{K_c}}\right)D_{\alpha,1}}\psi_{\lambda \alpha\beta}^\text{rest}(x)
\, e^{i\frac{\pi}{\sqrt{M}} \left( \lambda \frac{\sqrt{K_c}}{2}\left[{\tilde \phi}^+_{\alpha,b=1}(x)-{\tilde \phi}^+_{\alpha,b=1}(-x)\right]-\frac{1}{2\sqrt{K_c}}\left[{\tilde \phi}^+_{\alpha,b=1}(x)+{\tilde \phi}^+_{\alpha,b=1}(-x)\right]\right)}.
\end{equation}
\end{widetext}

\section{Elimination of dot-lead coupling}\label{can_trans}

The dot-lead interaction $H_\text{exc}$ in Eq.~(\ref{Hcoul}) can be eliminated by a canonical transformation as first noted by Schotte and Schotte.\cite{Schotte69} Using Eqs.~(\ref{Hcoul}), (\ref{rhodef}) and (\ref{constraintYY}) gives 
\begin{align}
H_\text{exc} ~=~ & \frac{\sqrt{K_cM}}{2} \sum_\alpha \int^\alpha dx \, W_\alpha(x)
\nonumber \\
& \times \, \left[ \nabla {\tilde \phi}_{\alpha,b=1}^+(x)-\nabla {\tilde \phi}_{\alpha,b=1}^+(-x)\right] n_d.
\label{Hexc1}
\end{align}
We now formally extend the integral over the full translational-invariant reference system by defining $W_\alpha(-x)=W_\alpha(x)$ (recall we have shifted the origin of coordinates in lead $\alpha$ to $x_\alpha$) obtaining
\begin{equation}
\\
H_\text{exc} = \frac{\sqrt{K_cM}}{2} \sum_\alpha \int dx \, W_\alpha(x) \, \nabla {\tilde \phi}_{\alpha,1}^+(x) \, n_d.
\label{Hexc2}
\end{equation}
Examination of Eqs.~(\ref{Hamiltonian3}) and (\ref{Hexc1}) then shows that the dot-lead coupling may be removed by a shift $\nabla {\tilde \phi^+_{\alpha,1}}(x)\rightarrow \nabla {\tilde \phi^+_{\alpha,1}}(x)-n_d Z_\alpha(x)$ with 
\begin{equation}
Z_\alpha(x)=\frac{\sqrt{K_cM} \, W_\alpha(x)}{\pi v_c}.
\label{Zdef}
\end{equation}

After this shift the Hamiltonian retains its original form, except that $H_\text{exc}$ has been eliminated and the dot parameters $\varepsilon_d$ and $U$ are shifted to $\varepsilon_d \rightarrow {\tilde \varepsilon_d}=\varepsilon_d-\Delta$ and $U \rightarrow {\tilde U}=U-2\Delta$
with the polaron shift $\Delta$ defined by
\begin{equation}
\Delta = \sum_\alpha\frac{K_cM}{4v_c\pi}\int dx \left[W_\alpha(x)\right]^2.
\label{eren}
\end{equation}
Observe that the magnitude of the renormalizations depends on the magnitude of $W_\alpha$, which in turn may reasonably be expected to vary with the distance between the dot and lead $\alpha$.

The commutation relation Eq.~(\ref{chiralcr}) shows that the momentum conjugate to $\nabla {\tilde \phi}^\pm_{\alpha,b=1}$ is $-\pi {\tilde \phi}^\pm_{\alpha,b=1}/2$ so that the shift is effected by the canonical transformation ${\cal O}\rightarrow e^{iS}{\cal O}e^{-iS}$ with
\begin{equation}
S = n_d \sum_\alpha \frac{\pi}{2}\int dx Z_\alpha(x) {\tilde \phi}_{\alpha,b=1}^+(x).
\label{Sdef}
\end{equation}
Under this transformation the hybridization term Eq.~(\ref{Hmix}) becomes
\begin{align}\label{transfhyb} 
e^{iS}H_\text{mix} e^{-iS} & ~=~ \sum_\alpha \int dx \, {\mathcal T}_\alpha(x) \, d^\dagger_{\sigma} \, e^{i\frac{\pi n_d}{4}B_{\alpha}(x)} 
\nonumber \\
\times \, & e^{i \frac{\pi}{2} \sum_{\alpha'} \int dx' Z_{\alpha'}(x') {\tilde \phi}_{\alpha',b=1}^+(x')} \, \psi_{\lambda \alpha \beta}(x).
\end{align}
Here the first factor in the second line comes from transforming the operator $d_{\sigma}^\dagger$ and the second from transforming the lead operator. The constant $B_\alpha(x)$ is given by
\begin{align}
B_\alpha(x) & = \lambda \sqrt{\frac{K_c}{M}}\int dx' \ \text{sgn}(x') \left[Z_\alpha(x'+x)-Z_\alpha(x'-x)\right]
\nonumber \\
- & \frac{1}{\sqrt{K_cM}}\int dx' \ \text{sgn}(x') \left[Z_\alpha(x'+x)+Z_\alpha(x'-x)\right].
\label{Bdef}
\end{align}
We shall be interested in $|x|$ small compared to the range over which $Z_\alpha$ is nonvanishing, in which case because $W_\alpha$ is defined as an even function $B_\alpha$ will be negligible.

\section{Excitonic interaction and dot-lead hybridization}\label{Electroniccorrelations}

The transformations introduced in Sec.~\ref{can_trans} remove the explicit excitonic dot-lead coupling from the Hamiltonian, at the expense of adding operator content to the hybridization. The resulting model can (to our knowledge) not be exactly solved, but it can be studied by perturbation theory in the dot-lead hybridization. We present here an investigation of the lowest nontrivial order in perturbation theory, which reveals the essential physics introduced by the excitonic dot-lead interaction. The methods introduced in Refs.~[\onlinecite{Werner07,Werner10}] can be used to extend the calculation numerically to all orders. 

In a perturbative analysis of Eq.~(\ref{transfhyb}) the crucial quantities are expectation values of the form
\begin{equation}\label{definecorrelator}
F_\alpha(t)= \sum_{\lambda,\lambda'=\pm} \left\langle 
\xi^\dagger_{\lambda \alpha \beta}(t) \xi_{\lambda' \alpha \beta}(0)
\right\rangle
\end{equation}
with
\begin{align}\label{definecorrelator2}
\xi_{\lambda \alpha \beta}& ~=~ U_{\alpha}\int dx \, {\mathcal T}_\alpha(x) \, e^{i\frac{\pi n_d}{4}B_\alpha(x)} \nonumber \\
& \times e^{i \frac{\pi}{2} \sum_{\alpha'} \int dx' Z_{\alpha'}(x') {\tilde \phi}_{\alpha',b=1}^+(x')} \, \psi_{\lambda \alpha \beta}(x).
\end{align}
Equation~(\ref{definecorrelator}) is evaluated by using Eq.~(\ref{psidef1}) to express the fermion operator in terms of chiral boson fields and then using standard results of bosonization. To avoid inessential complications we specialize to the physically most relevant case of short-ranged dot-lead interactions. In this case we may neglect the $B_\alpha(x)$ term in Eq.~(\ref{definecorrelator2}). We will also assume that (as in the nanotube case) all of the non-charge channels are characterized by very weak interactions, so that we may assume that in all channels except the charge channel we have free-fermion correlations. In the long-time limit we then obtain (including for later convenience a nonvanishing temperature $T$)
\begin{equation}
F_\alpha(t) \simeq \frac{1}{ \tau_\alpha} \left( \frac{v_F}{v_c} \right)^{\frac{1}{M}} F_0(t) \, e^{\Phi_\alpha(t)} \, e^{-i\mu_\alpha t} 
\label{Fdef}
\end{equation}
with $F_0(t)= \pi \eta T/ [v_F \sinh (\pi T t) ]$ the free-fermion correlator,
\begin{align}\label{definebarerate}
\frac{1}{\tau_\alpha} ~\equiv~ & 2\pi
\int dx \, dx' \, {\mathcal T}^*_\alpha(x) \, {\mathcal T}_\alpha(x') \nonumber \\
& ~ \times 4 \sin (k_F (x-x_\alpha)) \sin (k_F (x'-x_\alpha))
\end{align} 
proportional to the bare tunneling rate, and 
\begin{align}
\Phi_\alpha(t) = & \frac{1}{M} \left( 1- \left[ \frac{(1-Z_\alpha)^2}{K_c} + \frac{Z_{\bar{\alpha}}^2}{K_c} \right] \right) \nonumber \\
& \times \, \log \left( \frac{i v_{c}\Lambda}{\pi T} \sinh ( \pi T t ) \right),
\label{Phidef}
\end{align}
where we have defined $\bar{\alpha}= L$ if $\alpha=R$ ($\bar{\alpha}= R$ if $\alpha=L$) and $Z_\alpha \equiv \int dx \, Z_\alpha(x)$. Here $\Lambda$ is the momentum cutoff of the Luttinger-liquid behavior; in the nanotube case it is in essence the tube diameter, which is the length scale on which the Coulomb interaction is cut off. 

The integral in Eq.~(\ref{definebarerate}) reflects the vanishing of the lead fermion operator precisely at the edge of the lead, $x_\alpha$, and may give rise to a strong dependence of electronic correlations on the position of the quantum dot (relative to the leads) but need not be evaluated precisely for our subsequent considerations.

The main physical content of Eq.~(\ref{Phidef}) is that the function $F_\alpha(t)$ is renormalized from the free-fermion behavior $\propto [\sinh ( \pi T t )]^{-1}$ to $[\sinh ( \pi T t )]^{-Y_\alpha}$ with
\begin{equation}\label{defineY}
Y_\alpha = 1 - \frac{1}{M} \left[ 1- \frac{(1-Z_\alpha)^2}{K_c} - \frac{Z_{\bar{\alpha}}^2}{K_c} \right].
\end{equation}
Differences from $Y_\alpha=1$ corresponds to changes from the free-fermion situation. $Y_\alpha>1$ corresponds to a suppression of tunneling and $Y_\alpha<1$ to an enhancement.

In the absence of the excitonic coupling ($Z_\alpha=Z_{\bar{\alpha}}=0$) we obtain 
$Y_\alpha=1+(1-K_c)/K_cM$, the standard result for tunneling into the boundary of a 
Luttinger liquid with only density correlations.\cite{Kane92,Fabrizio97,Furusaki97}
The excitonic coupling has two effects. The $(1-Z_\alpha)^2/K_c$ term leads to a weakening of the charge-channel renormalization and thus to an enhancement of tunneling, similar to that found in our previous work.\cite{Elste} However, the term $Z_{\bar{\alpha}}^2/K_c$, which did not occur in our previous work, leads to a strengthening of the charge-channel renormalization and hence to a suppression of tunneling. This term arises because tunneling into one lead changes the dot charge density. The excitonic interaction means that this change in the dot density causes an orthogonality catastrophe in the other lead, suppressing electronic tunneling. The total renormalization is governed by a competition between the two effects, and is thus controlled by the relative sizes of $Z_\alpha$ in the two leads as well as by $K_c$. 

If the situation is symmetric and the interactions are of Coulombic origin with a reasonably large screening length, then as shown in Ref.~\onlinecite{Elste} one has $Z_\alpha=1-K_c^2$ with $K_c<1$. In this case one obtains
\begin{equation}
Y_L = Y_R = 1 + \frac{1}{K_cM} \left[ 1-K_c - 2K_c^2 (1 - K_c^2) \right].
\end{equation}
Especially in the strong-interaction case ($K_c\ll 1$), the value of $Y_\alpha$ is changed relatively little from the $Z_\alpha=0$ value. For a nanotube with a screening length somewhat larger than the tube diameter, the considerations of Ref.~\onlinecite{Elste} imply $K_c$ on the order of $0.5\text{--}0.7$. In this circumstance, $Y_\alpha$ is $1.1\text{--}1.3$ in the absence of excitonic effects while in its presence $Y_\alpha$ is $0.9\text{--}1.1$.

On the other hand, a very asymmetric situation ($Z_L\simeq 1$, $Z_R\simeq 0$) would lead to $Y_L\simeq 1- 1/M<1$ and $Y_R\simeq 1+ (2/K_c-1)/M > 1$, so that tunneling into one of the leads is suppressed and the other is enhanced; however, it is difficult to realize a small $Y_\alpha$ even in the case of $M=1$ channel.

\section{Current-voltage characteristics}\label{Currentvoltagecharacteristics}

The assumption of weak coupling between the quantum dot and the leads
suggests a master-equation approach to study the nonequilibrium dynamics of the system.
The quantum dot with a spin-degenerate level is described by the diagonal density matrix 
\begin{equation}\label{rhodiag1}
\rho_d = \mathcal{P}_0 |0\rangle \langle 0| + \mathcal{P}_1 | 1 \rangle \langle 1| + \mathcal{P}_2 | 2 \rangle \langle 2|,
\end{equation}
where $\mathcal{P}_0$, $\mathcal{P}_1$, and $\mathcal{P}_2$ denote the occupation probabilities of the empty state $|0\rangle$, the singly-charged state $|1 \rangle $, and the doubly-charged state $|2 \rangle $. 

In the Markovian limit of sufficiently slow dot dynamics, inserting Eq.~(\ref{rhodiag1}) into the von Neumann equation and expanding to lowest order in the hybridization yields the following set of master equations
\begin{align}
\dot{\mathcal{P}}_0 &~=~ \sum_{\alpha} \left[ \mathcal{P}_1 \mathcal{R}^{\alpha}_{1 \rightarrow 0} - 2\mathcal{P}_0 \mathcal{R}^{\alpha}_{0 \rightarrow 1} \right], \label{rate-equation1} \\
\dot{\mathcal{P}}_1 &~=~ \sum_{\alpha} \left[ 2 \mathcal{P}_0 \mathcal{R}^{\alpha}_{0 \rightarrow 1} + 2\mathcal{P}_2 \mathcal{R}^{\alpha}_{2 \rightarrow 1} - \mathcal{P}_1 \left( \mathcal{R}^{\alpha}_{1 \rightarrow 0} + \mathcal{R}^{\alpha}_{1 \rightarrow 2} \right) \right], \label{rate-equation2} \\
\dot{\mathcal{P}}_2 &~=~ \sum_{\alpha} \left[ \mathcal{P}_1 \mathcal{R}^{\alpha}_{1 \rightarrow 2} - 2\mathcal{P}_2 \mathcal{R}^{\alpha}_{2 \rightarrow 1} \right], \label{rate-equation3} 
\end{align}
with the tunneling rates 
\begin{align}
\mathcal{R}^{\alpha}_{0 \rightarrow 1} & = 2 \, \text{Re} \int_0^\infty d\tau F_\alpha(\tau) \, e^{-i\tilde{\varepsilon}_d\tau}, \\
\mathcal{R}^{\alpha}_{1 \rightarrow 2} & = 2 \, \text{Re} \int_0^\infty d\tau F_\alpha(\tau) \, e^{-i(\tilde{\varepsilon}_d+\tilde{U})\tau}.
\end{align}

\begin{figure}[!t]
\begin{center}
$\begin{array}{c}
\textbf{(a)} \\ \\
\includegraphics[width=8.0cm,angle=0]{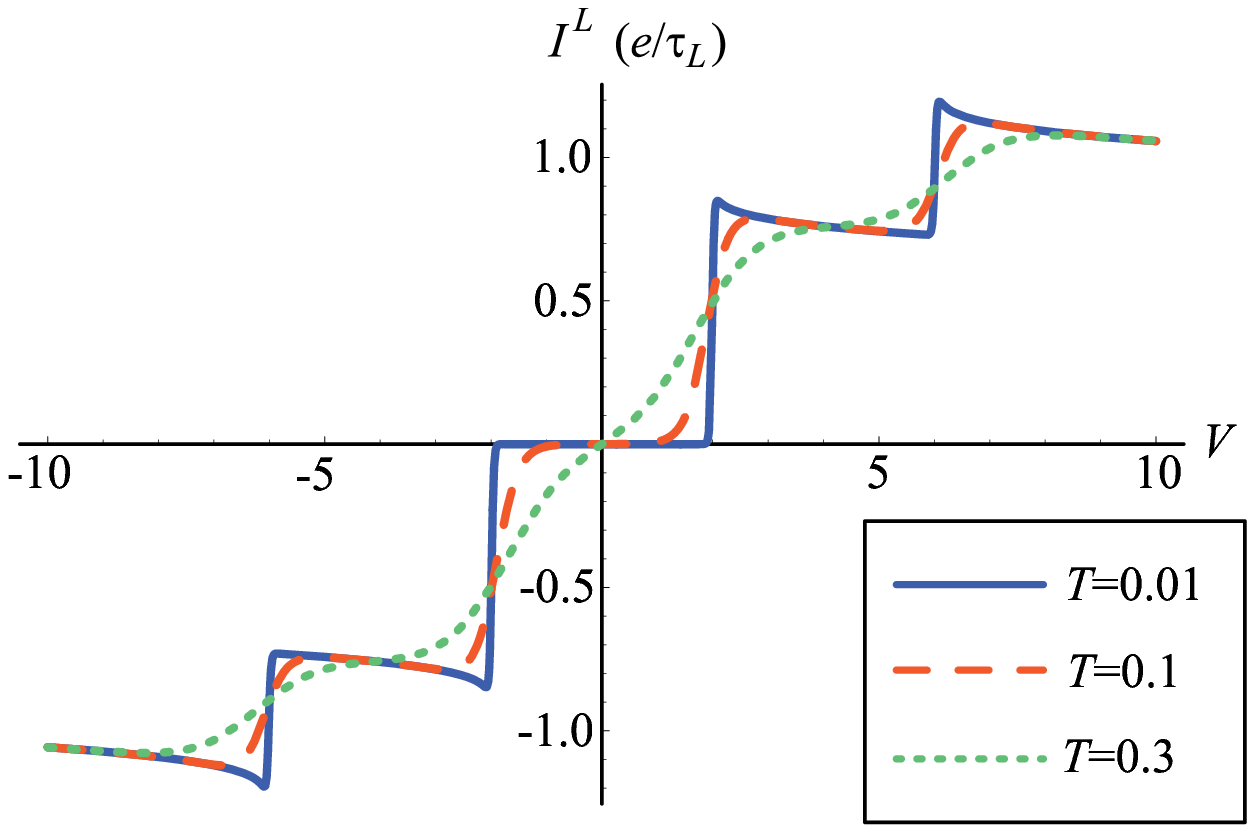} \\
\textbf{(b)} \\ \\
\includegraphics[width=8.0cm,angle=0]{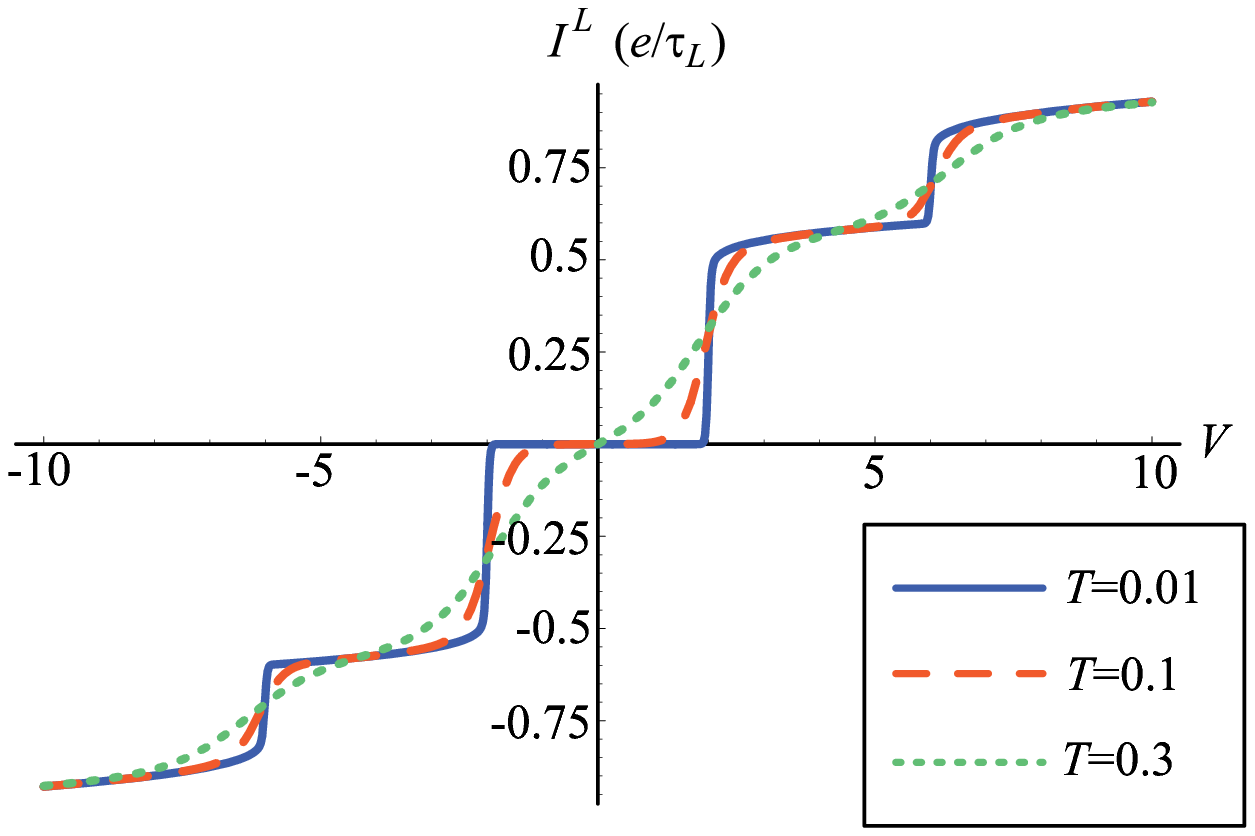}
\end{array}$
\caption{Current-voltage characteristics of a broken Luttinger liquid (a) for $Y=0.9$ (representative of the presence of the excitonic interaction in a symmetric junction with nanotube leads) and (b) for $Y=1.1$ (representative of a dot with nanotube leads in the absence of the excitonic interaction), and different temperatures. We show results for a symmetrically applied bias voltage $eV=\mu_L-\mu_R$ with the gate voltage set to zero. We assume that the renormalized and bare dot energies ${\tilde \varepsilon_d}$ and $\varepsilon_d$ are positive. Because our interest here is in the form of the threshold behavior we choose units such that in each panel the bias voltages $V$ and thermal energies $T$ are given in units of the corresponding onsite energy appropriate to that panel. 
The local Coulomb interaction is assumed to be twice as large as the onsite energy in both panels.}\label{IVcurves} 
\end{center}
\end{figure}

The transition probabilities for the emission and absorption of an electron are related by
$\mathcal{R}^{\alpha}_{0 \rightarrow 1}(\tilde{\varepsilon}_d-\mu_\alpha)=\mathcal{R}^{\alpha}_{1 \rightarrow 0}(-\tilde{\varepsilon}_d+\mu_\alpha)$ and 
$\mathcal{R}^{\alpha}_{1 \rightarrow 2}(\tilde{\varepsilon}_d+\tilde{U}-\mu_\alpha)=\mathcal{R}^{\alpha}_{2 \rightarrow 1}(-\tilde{\varepsilon}_d-\tilde{U}+\mu_\alpha)$, respectively. Using Eq.~(\ref{Fdef}) we find 
\begin{align}\label{rates_finiteT1}
\mathcal{R}^{\alpha}_{0 \rightarrow 1} & \propto
\frac{e^{-\frac{\tilde{\varepsilon}_d-\mu_\alpha}{2T}}}{2\pi\tau_\alpha} \left(\frac{2\pi T}{v_c\Lambda}\right)^{Y_\alpha-1} \, 
\frac{\left| \Gamma \left( \frac{Y_\alpha}{2} + i \frac{\tilde{\varepsilon}_d-\mu_\alpha}{2\pi T} \right) \right|^2}{\Gamma(Y_\alpha)}, \\ \label{rates_finiteT2}
\mathcal{R}^{\alpha}_{1 \rightarrow 2} & \propto
\frac{e^{-\frac{\tilde{\varepsilon}_d+\tilde{U}-\mu_\alpha}{2T}}}{2\pi\tau_\alpha} \left(\frac{2\pi T}{v_c\Lambda}\right)^{Y_\alpha-1} \, 
\frac{\left| \Gamma \left( \frac{Y_\alpha}{2} + i \frac{\tilde{\varepsilon}_d+\tilde{U}-\mu_\alpha}{2\pi T} \right) \right|^2}{\Gamma(Y_\alpha)}.
\end{align}
The rates in Eqs.~(\ref{rates_finiteT1})---(\ref{rates_finiteT2}) obey the detailed-balance condition, because each lead is individually in equilibrium. 

\begin{figure}[!!t]
\begin{center}
\includegraphics[width=8.0cm,angle=0]{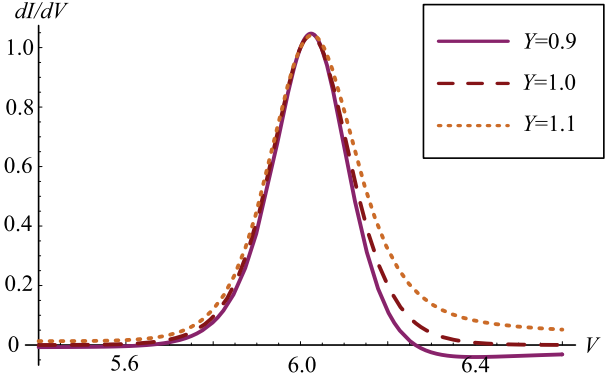}
\caption{Differential conductance $dI/dV$ (in units of its maximum value) as a function of bias voltage $V$
for different values of $Y$ calculated for voltages in the neighborhood of the higher bias step shown in Fig.~\ref{IVcurves} at temperature $T=\tilde{\varepsilon}_d/30$ and other parameters identical to those used in Fig.~\ref{IVcurves}. }\label{dIdV} 
\end{center}
\end{figure}

The operator for the tunneling current through lead $\alpha$ is
\begin{equation}
I_{\alpha} = -i e \sum_{a \sigma} \int^\alpha dx \left[ \mathcal{T}_\alpha(x) d_\sigma^{\dagger} \psi_{\alpha a \sigma}(x) - \mathcal{T}_\alpha^*(x) \psi^\dagger_{\alpha a \sigma}(x) d_\sigma \right].
\end{equation}
Solving Eqs.~(\ref{rate-equation1})---(\ref{rate-equation3}) in steady state ($\dot{\mathcal{P}}_i=0$)
then gives the steady-state current
\begin{align}\label{analytic_ex}
\langle I_{\alpha} \rangle & ~=~ 
e\,\frac{\mathcal{R}_{2 \rightarrow 1}}{R} \left( \mathcal{R}^{\alpha}_{0 \rightarrow 1} \mathcal{R}^{\bar{\alpha}}_{1 \rightarrow 0} - \mathcal{R}^{\alpha}_{1 \rightarrow 0} \mathcal{R}^{\bar{\alpha}}_{0 \rightarrow 1} \right) \nonumber \\
&~+e\,\frac{\mathcal{R}_{0 \rightarrow 1}}{R}
\left( \mathcal{R}^{\alpha}_{1 \rightarrow 2} \mathcal{R}^{\bar{\alpha}}_{2 \rightarrow 1} - \mathcal{R}^{\alpha}_{2 \rightarrow 1} \mathcal{R}^{\bar{\alpha}}_{1 \rightarrow 2} \right)
\end{align}
with $\mathcal{R}_{n \rightarrow m}\equiv\sum_\alpha \mathcal{R}^\alpha_{n \rightarrow m}$ and 
\begin{equation}
R= \mathcal{R}_{1 \rightarrow 0} \mathcal{R}_{2 \rightarrow 1} +
2 \mathcal{R}_{0 \rightarrow 1} \mathcal{R}_{2 \rightarrow 1}
+\mathcal{R}_{0 \rightarrow 1} \mathcal{R}_{1 \rightarrow 2}.
\end{equation}
Thus we see explicitly that the steady-state current is proportional to the product of the hybridization to the left and right lead and changes when the chemical potential in either lead lines up with one of the dot levels. In the asymmetric situation different terms come in to resonance at different bias voltages, leading to a diversity of resonance behaviors. 

In the low-temperature limit, the current-voltage characteristics obey a power law close to the Coulomb-blockade thresholds, $eV=\pm 2\tilde{\varepsilon}_d$. Here the electronic correlation function is
$F_\alpha \propto 1/t^{Y_\alpha}$, which gives rise to tunneling rates of the form
\begin{equation}
\mathcal{R}^{\alpha}_{0 \rightarrow 1} \propto
\frac{1}{\tau_\alpha} \, \left(\frac{|\tilde{\varepsilon}_d-\mu_\alpha|}{v_c\Lambda}\right)^{Y_\alpha-1} \, \theta\left( \mu_\alpha-\tilde{\varepsilon}_d \right).
\label{T0rate}
\end{equation}

Numerical results for the current-voltage characteristics for a symmetric junction (equal hybridization and excitonic couplings to left and right lead) are presented in Fig.~\ref{IVcurves}, for exponent values representative of a dot coupled to carbon nanotube leads with (upper panel) and without (lower panel) excitonic couplings. The steps in the current are associated with the lead Fermi level coming to resonance with the $n_d=0$ to $n_d=1$ energy difference (lower bias feature) and the $n_d=1$ to $n_d=2$ energy difference (higher bias feature). The symmetry of the situation implies that the steps are the same for positive and negative bias. We see that in the excitonic case (upper panel) the conductance is enhanced at the threshold, while in the absence of excitonic coupling the conductance is suppressed at the threshold. As is seen more clearly in Fig.~\ref{dIdV} the excitonic enhancement is accompanied by a region of negative differential conductance.

\begin{figure}[!t]
\begin{center}
$\begin{array}{c}
\textbf{(a)}~~~~~~~~~~~~~~~~~~~~~~~~~~~~\textbf{(b)}~~~~~~~~ \\
\includegraphics[width=8.5cm,angle=0]{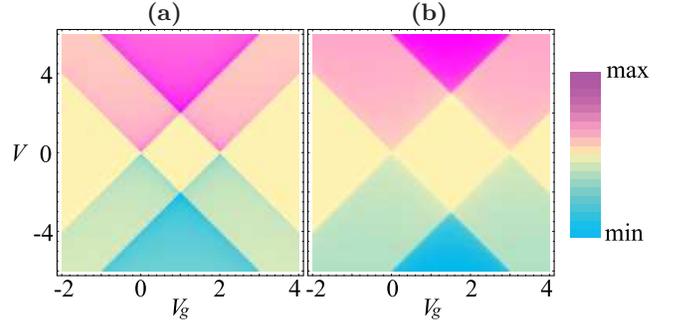}
\end{array}$
\caption{Color-scale plots of the current as a function of bias voltage $V$
and gate voltage $V_g$ 
(a) for $Y=0.9$ (representing the presence of the excitonic interaction) and
(b) for $Y=1.1$ (representing the absence of the excitonic interaction).
Voltages are given in units of the onsite energy. 
We assume $T=\varepsilon_d/100$, $U=3\varepsilon_d$, and $\Delta=\varepsilon_d/2$ so that $\tilde{U}=2\varepsilon_d$.}\label{Coulombdiamonds}
\end{center}
\end{figure}

Color-scale plots of the current as a function of bias voltage $V=(\mu_L-\mu_R)/e$ and gate voltage $V_g=-\varepsilon_d/e$ are shown in Fig.~\ref{Coulombdiamonds}. The excitonic interaction leads to a polaron shift, $\Delta$, which renormalizes the dot charing energy, $U \rightarrow \tilde{U}=U-2\Delta$, and thus reduces the size of the Coulomb diamonds.

\begin{figure}[!t]
\begin{center}
$\begin{array}{c}
\textbf{(a)} \\ \\
\includegraphics[width=8.0cm,angle=0]{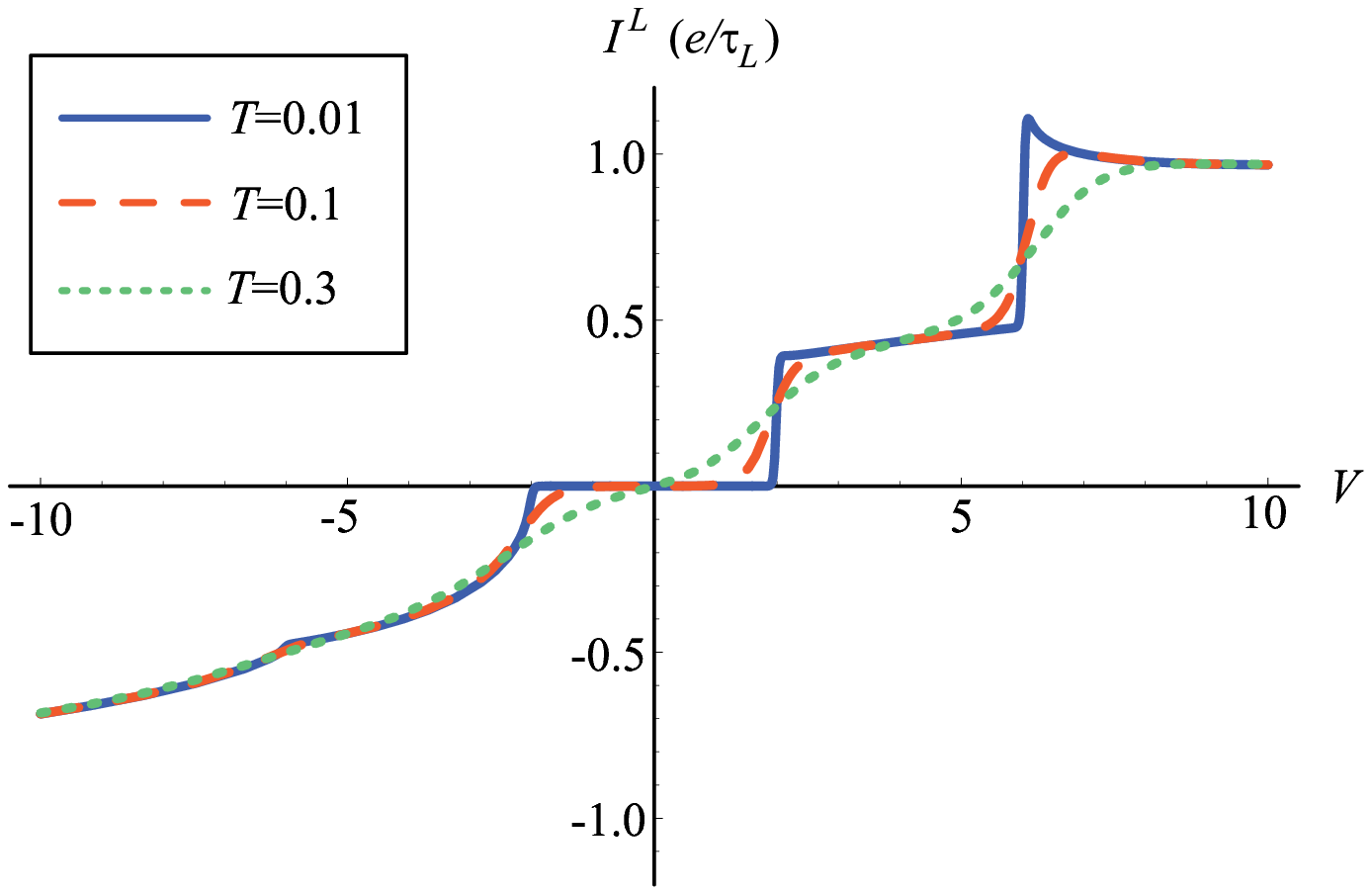} \\
\textbf{(b)} \\ \\
\includegraphics[width=8.0cm,angle=0]{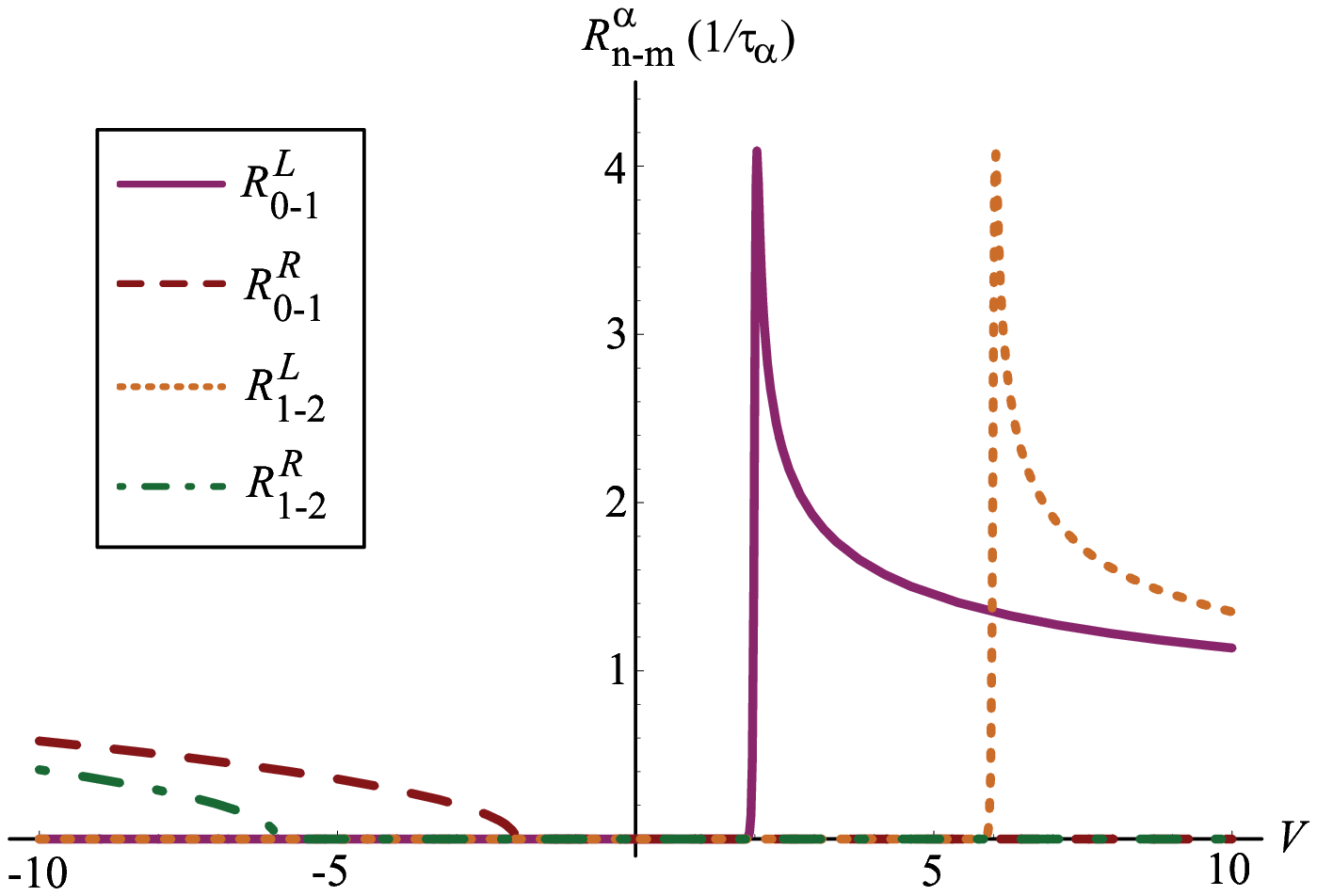}
\end{array}$
\caption{(a) Current-voltage characteristics and (b) transition rates for asymmetric excitonic dot-lead couplings, $Y_L=0.75$ and $Y_R=1.5$. 
We assume $T=\tilde{\varepsilon}_d/100$, $v_c\Lambda=15{\tilde \varepsilon_d}$ and as in Fig.~\ref{IVcurves} choose ${\tilde U}=2{\tilde \varepsilon}_d$. Voltages are given in units of the onsite energy.}\label{current_asym} 
\end{center}
\end{figure}

As discussed in Sec.~\ref{Electroniccorrelations}, asymmetric excitonic dot-lead couplings change the behavior. In Fig.~\ref{current_asym} we demonstrate this behavior, presenting results calculated for parameters appropriate to a nanotube ($M=4$ channels and $K_c\simeq 0.7$) but with an excitonic interaction only to the left lead, so that $Z_L \simeq 1$ and $Z_R \simeq 0$, implying $Y_L\simeq 0.75$ and $Y_R\simeq 1.5$. In this case tunneling to the left lead is enhanced and tunneling to the right lead is suppressed. To understand the consequences for the $I$--$V$ characteristics we refer to Eq.~(\ref{analytic_ex}) for the steady-state current. Near the lowest threshold we may neglect the 1$\rightarrow$2 transitions and simplify the expression to $\langle I_L \rangle = 2e ( \mathcal{R}^L_{0 \rightarrow 1} \mathcal{R}^R_{1 \rightarrow 0}-\mathcal{R}^L_{1 \rightarrow 0} \mathcal{R}^R_{0 \rightarrow 1})/(2\mathcal{R}_{0 \rightarrow 1}+\mathcal{R}_{1 \rightarrow 0})$. We see that when the bias voltage reaches the positive threshold $eV=+2\tilde{\varepsilon}_d$, only the first term contributes because $\mathcal{R}^R_{0 \rightarrow 1}\simeq 0$ in this case. Since the rate $\mathcal{R}^R_{1 \rightarrow 0}$ is a smooth function at $eV=+2\tilde{\varepsilon}_d$ and the denominator has no singularities,
the current is essentially proportional to $e\mathcal{R}^L_{0 \rightarrow 1}$. However, when the bias voltage reaches the negative threshold $eV=-2\tilde{\varepsilon}_d$, only the second term contributes because $\mathcal{R}^L_{0 \rightarrow 1}\simeq 0$ in that case. Accordingly the current is asymmetric and shows step-like features at positive bias voltages but a suppressed onset at negative bias voltages. In the asymmetric case the ``Luttinger'' renormalization of the tunneling amplitude is different for tunneling into the two leads [cf.~Eq.~(\ref{T0rate})], also contributing to the asymmetry.

\section{Conclusions}\label{Conclusions}

In summary, we have investigated the effect of a Coulombic dot-lead coupling on transport through a quantum dot coupled to two semi-infinite Luttinger-liquid leads. The electronic tunneling has been described within a master-equation approach that treats the dot-lead hybridization to lowest nonvanishing order. We have found that the excitonic dot-lead interaction may enhance transport in the vicinity of the Coulomb-blockade threshold, which is in contrast to the power-law suppression of the electronic tunneling if this interaction is not included. However, the effects are in general less pronounced than for the side-coupled situation considered in our previous work,\cite{Elste} because in the present two-lead situation both the excitonic and orthogonality effects are present.

The paper raises several questions for future research. A treatment of the electronic tunneling to all orders in the hybridization would be desirable. Moreover, it would be interesting to study the effect of the Coulombic dot-lead coupling on transport in the Kondo regime, where electronic tunneling is dominated by dot-lead exchange scattering processes. Finally, the consequences of an interaction between the two leads should be explored.

\acknowledgments
AJM acknowledges support from the National \mbox{Science} Foundation under grant DMR-0705847. FE acknowledges support from the Deut\-sche For\-schungs\-ge\-mein\-schaft.

\end{document}